\documentclass[%
 aip,
 pop,
 amsmath,amssymb,
 reprint,%
]{revtex4-1}

\usepackage{graphicx}
\usepackage{dcolumn}
\usepackage{bm}

\usepackage[utf8]{inputenc}
\usepackage[T1]{fontenc}
\usepackage{mathptmx}

\newcommand{\beq}{\begin{equation}}
\newcommand{\eeq}{\end{equation}}
\newcommand{\vu}{\vec{u}}
\newcommand{\vb}{\vec{b}}

\let\oldhat\hat
\renewcommand{\vec}[1]{{\bm{#1}}}
\renewcommand{\hat}[1]{{\oldhat{\bm{#1}}}}

\begin{document}

\preprint{AIP/123-QED}

\title[Chaotic behavior of Eulerian magnetohydrodynamic turbulence]{Chaotic behavior of Eulerian magnetohydrodynamic turbulence}

\author{Richard D. J. G. Ho}
 \email{richard.ho@ed.ac.uk}
\author{Arjun Berera}
 \email{ab@ph.ed.ac.uk}
\author{Daniel Clark} 
\affiliation{SUPA, School of Physics and Astronomy, University of Edinburgh, \\
JCMB, King's Buildings, Peter Guthrie Tait Road, Edinburgh EH9 3FD, United Kingdom
}%

\date{\today}

\begin{abstract}
We study the chaotic properties of a turbulent 
conducting fluid using direct numerical simulation
in the Eulerian frame.
The maximal Lyapunov exponent is measured for simulations with varying Reynolds number
and magnetic Prandtl number.
We extend the Ruelle theory of hydrodynamic turbulence to magnetohydrodynamic turbulence
as a working hypothesis and find broad agreement with results.
In other simulations we introduce magnetic helicity and these simulations
show a diminution of chaos, 
which is expected to be eliminated at maximum helicity.
We also find that the difference between two initially close fields grows linearly
at late times, which was also recently found in hydrodynamics.
This linear growth rate is found to be dependent on the dissipation rate of the relevant field.
We discuss the important consequences this linear growth has on predictability.
We infer that the chaos in the system is totally dominated by the velocity field and
connect this work to real magnetic systems such as solar weather and confined plasmas.
\end{abstract}

\maketitle

\section{\label{sec:level1}Introduction}

Turbulence displays chaotic dynamics. A small change in initial conditions will
result in a large difference in the state at later times, with this difference
growing exponentially.
This exponential growth of error puts limits on the predictability of the system,
and understanding these limits is important for forecasting the evolution of
fluids governed by turbulence.
Turbulence, or its underlying equations, is interesting from a
dynamical systems point of view. In this context, it is just another dynamical
system of which we wish to know the chaotic properties.

In homogeneous isotropic turbulence (HIT), recent results have shown a relationship between
the Reynolds number, Re, and the maximal Lyapunov exponent, $\lambda$, for chaos in an 
Eulerian (considering the difference between two fields) description \cite{Berera2018,Boffetta2017}
consistent with theoretical predictions by Ruelle \cite{Ruelle1979,Crisanti1993}.
In these simulations, two initially close fields were evolved concurrently and their
difference quantified.
Surprisingly, a limit was found on the growth of this difference which 
is proportional to the dissipation rate.

This paper presents analysis of a set of simulations of magnetohydrodynamic (MHD)
turbulence, also known as hydromagnetic turbulence.
This is the first analysis of chaos in MHD turbulence in an Eulerian context.
Within this first analysis, we produce a dataset 
for chaotic behavior in MHD in the range of Prandtl number that is
comparable to all the simulation data to date on MHD spectra.
The analysis of these simulations and presentation of the data from them is the main work of this paper.
We develop a working hypothesis that 
the level of chaos primarily depends on quantities dependent on the velocity field only.
This hypothesis is found to show reasonable consistency with the results from our 
simulations.
There is room for further theoretical consideration but that is beyond the scope
of this paper which is only focussed on the measurements of the Lyapunov exponents.
Our simulations also show a growth limit analogous to that in hydrodynamics,
but for MHD the value of this limit differs for the magnetic versus velocity fields.

The evolution of an uncharged fluid is described by the Navier-Stokes equation, whilst
the evolution of an electrically conducting fluid is described by the MHD equations.
The incompressible MHD equations are
\begin{align}
\label{eq:MHD}
\partial_t \vec{u} &= - \frac{1}{\rho}\nabla P -(\vec{u}\cdot \nabla)\vec{u} + \nu \Delta \vec{u} + \frac{1}{\rho}(\nabla \times \vec{b}) \times \vec{b} \ , \\
\partial_t \vec{b} &= (\vec{b} \cdot \nabla) \vec{u} - (\vec{u} \cdot \nabla) \vec{b} + \eta \Delta \vec{b} \ , \ 
\nabla \cdot \vec{u} = \nabla \cdot \vec{b} = 0 \ ,
\end{align}
with velocity field $\vec{u}$, magnetic field $\vec{b}$, density $\rho$, 
pressure $P$, viscosity $\nu$, and magnetic diffusivity $\eta$.
Like the Navier-Stokes equations which they modify, they also display turbulence \cite{Biskamp2003}.

In the equations above, the incompressible assumption is made.
In many real world applications, such as at
laboratories and space plasma systems, the incompressible assumption is made 
merely for convenience, and 
it is known that compressible effects can be of importance.
As well, there may be complications from the reduced MHD limit, where there are
very large guide magnetic fields \cite{Kadomtsev1974,Strauss1976}.
The velocity perpendicular to the guide field is incompressible, but the parallel component can 
be far from incompressible, and in fact sound waves can play an important role \cite{Zank1993}.
However, as the simulations analysed in this paper use the incompressible MHD equations, they
are presented here in the above form.

The MHD equations contain three ideal inviscid invariants which should hold within the
inertial range.
These are total energy, magnetic helicity, and magnetic cross helicity.
The magnetic helicity, although not positive definite, can undergo an inverse transfer
from the small to large length scales \cite{Brandenburg2001,Alexakis2006}. This transfer
itself is considered to not be a proper cascade since it vanishes in the limit of infinite magnetic
Reynolds number \cite{Alexakis2018}.
Whilst these are the invariants for HIT, other invariants, which can be important,
may exist and depend on the geometric configuration.


The literature investigating the intersection of MHD and chaos is sparse.
The relative dispersion of charged particles is predicted to grow exponentially for
intermediate times when they are initially close \cite{Misguich1982,Misguich1987}.
The technique of extracting a Lyapunov-exponent spectrum from a time series \cite{Wolf1985}
has been applied to experimental data from an undriven plasma system \cite{Huang1994}.
Use of this technique reveals a transition from quasiperiodicity to chaos and allows
the evaluation of the Lyapunov spectrum.
In solar physics, a low dimensional attractor is suggested to be responsible for the data
seen in pulsation events of solar radio emissions \cite{Kurths1986}.
Shell models of turbulence applied to MHD with Pr $\sim 1$ have found \cite{Grappin1986} that the
maximal Lyapunov exponent obeys $\lambda \sim \nu^{-1/2}$. 
Later results for turbulent models of Navier-Stokes found a similar scaling \cite{Yamada1987},
which is roughly the scaling found in DNS \cite{Berera2018}.


MHD turbulence has a wide variety of applications, from turbulence in the
solar wind \cite{Tu1995,Goldstein1995},
accretion disks \cite{Balbus1998,MacLow2004}, 
and the interstellar medium \cite{Goldreich1995,MacLow1999}.
A lot of work in MHD turbulence has looked at magnetic reconnection
at small scales, where compression enters as an important
effect as well \cite{Biskamp1996,Yamada2010}.
The effect of chaos on these physical systems should be understood, and could
give new understanding of the dynamics.

The relationship between MHD and chaos is also 
interesting from the viewpoint of predictability.
Just as it is important to quantify the predictability of weather forecasts to 
understand the time horizons over which a prediction can be considered accurate, 
it is important to understand the predictability of forecasts where the 
governing equations are the MHD equations, such as in space weather \cite{Mirmomeni2009}, solar physics \cite{Karak2012},
and high latitude ground magnetic fields \cite{Weigel2003}.
For instance, much effort is put into understanding solar flares, a phenomena governed by
the MHD equations, because of the damage they can cause to artificial satellites and thus global communications \cite{Amari2014}.
Whilst HIT is an ideal description of turbulence, it nonetheless should represent the behaviour
of a turbulent system far from boundaries or at small scales.
The work here may have practical use in these situations but also in tokamak reactors and fusion research
where boundaries are present. More generally, establishing measures
of chaos in turbulent fluids provides another probe alongside spectra
for understanding the behavior of such complex systems.

The paper is organized as follows: Section \ref{section3} 
extends the Ruelle prediction for hydrodynamic turbulence to MHD. Section \ref{sectionDNS}
describes the code used and method for calculating Lyapunov exponents from the simulations.
Section \ref{section5} goes over the results of the simulations and Section \ref{sectionConclusion}
discusses implications of the results and application to other systems.

\subsection{\label{section3}Working hypothesis for $\lambda$ in MHD}

In a chaotic system, for two states which initially differ by 
separation $\delta_0$, 
this separation will grow as $\delta(t) \sim 
\delta_0 \exp(\lambda t)$, where $\lambda$ is the maximal Lyapunov exponent.
For fluid turbulence, this separation can either be particle positions within a Lagrangian
description, or a measure of the difference between two fields within an Eulerian description.
According to the theory of Ruelle \cite{Ruelle1979}, the maximal Lyapunov exponent for
Navier-Stokes turbulence is given by
\begin{align}
\lambda \sim \frac{1}{\tau} \ ,
\end{align}
where $\tau = \sqrt{\nu/\varepsilon_k}$ is the Kolmogorov microscale time and $\varepsilon_k$ the kinetic dissipation.
The argument used by Ruelle requires the existence of an inertial range in order to justify
the existence of a characteristic exponent 
which is dependent only on the dissipation.
Ruelle's arguments made no assumption about the frame of reference,
so are equally applicable to both the Eulerian and Lagrangian
descriptions.  The exact relation between the Eulerian and Lagrangian
descriptions for many quantities is unknown or very difficult to determine, 
but for application
to this work of Ruelle this does not seem to be a major issue.
In hydrodynamic turbulence this relation becomes
\begin{align}
\label{eq:alpha}
\lambda \sim \frac{1}{T_0} Re^\alpha \ ,
\end{align}
where $T_0 = L/u$ is the large eddy turnover time, Re $= uL/\nu$ is the Reynolds number,
$u$ is the rms velocity, and $L$ the integral length scale,  
The Kolmogorov theory predicts that $\alpha = 0.5$ \cite{Crisanti1993,Kolmogorov1941a}.
Intermittency corrections predict that $\alpha \lesssim 0.5$.
However, DNS results have shown that, in an Eulerian description, 
$\alpha \gtrsim 0.5$ \cite{Berera2018,Boffetta2017,Mohan2017}
whilst in a 
Lagrangian description $\alpha \lesssim 0.5$ \cite{Biferale2005}.
These DNS results also show a corresponding behaviour for $\lambda \tau$ 
which either rises (for $\alpha > 0.5$) or falls (for $\alpha < 0.5$) with Re.

We now look at how Ruelle's arguments can be extended to MHD.
For this, observe that in the MHD evolution equations, Eq. (\ref{eq:MHD}),
there is only a direct non-linearity for the velocity field itself.
In contrast, the magnetic field is only indirectly non-linear via
the velocity field.
As such, 
we predict that the chaos due to the velocity field is dominant over that for 
the magnetic field.
Thus, the Ruelle prediction, which relates the Lyapunov exponent to the 
smallest timescale \cite{Ruelle1979}, should only need to
be modified slightly.
We hypothesise that it depends on the smallest timescale that is itself
dependent only on velocity field quantities.
Thus we argue that the Ruelle prediction that $\lambda \sim \sqrt{\varepsilon_k / \nu}$, where
$\varepsilon_k$ is the kinetic dissipation, should also hold for MHD
in an Eulerian sense.
This prediction that $\lambda \sim \nu^{-1/2}$ is also backed up by the findings of 
shell models of turbulence
mentioned previously \cite{Grappin1986,Yamada1987}.

Although we expect the chaos to be dominated by the velocity field quantities, we
cannot rule out that the magnetic field could strongly affect the chaos.
Indeed, in Section \ref{sec:hel}, we find that an increase in magnetic helicity,
a quantity which only depends on the magnetic field, decreases the level of chaos
in the system.
A similar effect might happen if there were
a particularly strong alpha effect \cite{Brandenburg2005}.

In extending findings of Eulerian chaos in hydrodynamics to MHD, there are other complications that must be tested, such as 
whether $\lambda \tau$ has any dependence on Re or Pr = $\nu/\eta$, 
the magnetic Prandtl number.
Pr is known to have important effects on dissipation rates and the presence
of inverse spectral transfer \cite{McKay2018}.
This paper relies on this previous foundational work.


Results from DNS simulations suggest that the ratio of $\varepsilon_k$ and $\varepsilon_b$ 
(where $\varepsilon_b$ is the magnetic dissipation)
depends on 
the Prandtl number according to the relationship $\varepsilon_k/\varepsilon_b \sim$ Pr$^q$
\cite{Brandenburg2014,McKay2018}
where $q$ depends on the presence of helicity 
in the system with $q > 0$ and so $\varepsilon_k$ should
become totally dominant over $\varepsilon_b$.
In hydrodynamics the growth rate of error is limited by $\varepsilon$
\cite{Berera2018,Boffetta2017}. This may carry over to MHD, and the specific dependence on 
either $\varepsilon_k$ or $\varepsilon_b$ needs to be tested.
These are examined in Section \ref{section5}.

\section{Direct numerical simulation}
\label{sectionDNS}

We performed direct numerical simulations of forced HIT on the 
incompressible MHD equations using a fully de-aliased
pseudo-spectral code in a periodic cube of length $2\pi$
with unit density.
An external forcing function $\vec{f}$ was applied to maintain energy
in the system.
The code and forcing are fully described in \cite{YoffeThesis,LinkmannThesis,McKay2017} and summarized here.
The field is initialised with a set of random variables following a Gaussian distribution with zero mean.
The initial kinetic and magnetic energy spectra were
$E_{b,u} (k, t = 0) = A k^4 \exp(k^2 /(2k_0)^2),$
where $k_0 = 5$ is the peak wave number.

The primary forcing used was an adjustable helicity forcing,
fully described in \cite{McKay2017}.
In this forcing, a helical basis composed of eigenvectors of the curl operator, $\bm{e}_1$,
$\bm{e}_2$ is used. These are unit vectors which satisfy 
$i \bm{k} \times \bm{e}_1 = k \bm{e}_1$ and
$i \bm{k} \times \bm{e}_2 = - k \bm{e}_2$.
These basis vectors are constructed
from a unit vector which is randomised at each time step.
The forcing in Fourier space is $\hat{\bm{f}} (\bm{k},t) = 
A(\bm{k})\bm{e}_1(\bm{k},t) + B(\bm{k})\bm{e}_2(\bm{k},t)$ for
the forced wave numbers, $k_f \leq 2.5$.
The co-efficients $A(\bm{k})$ and $B(\bm{k})$ can be controlled to 
adjust the helicity of the forcing (either kinetic helicity if the velocity
field were forced, or magnetic helicity if the magnetic field were forced).

In all simulations, only the velocity field was forced
except those in Section \ref{sec:hel}.
The co-efficients $A(\bm{k})$ and $B(\bm{k})$ were chosen such that
no kinetic helicity was injected into the system. 
The magnetic helicity was roughly zero
although it was not directly controlled in these simulations
as described in \cite{McKay2017}.
In the simulations used in Section \ref{sec:hel}, only the magnetic field was
forced. When the magnetic field was forced,
$A(\bm{k})$ and $B(\bm{k})$ were chosen
in order to control the magnetic helicity.
The necessary benchmarking for the simulations 
has been done in \cite{McKay2018}.
This benchmarking included making sure that all simulations were fully resolved in both fields.
This was key in allowing greater confidence in the accuracy of the final results presented in this paper.

To measure the chaotic properties of the fields, the following
procedure was used.
After a statistically steady state was reached, a copy of evolved fields
$\bm{u}_1$ and $\bm{b}_1$ were made. At one time point, these fields were
perturbed by adding a white noise to them of size $\delta_0$,
creating fields $\bm{u}_2$ and $\bm{b}_2$.
This introduced a difference
between the fields, with both sets of fields then evolved independently.
The difference spectra $E_{ud}(k,t)$ and $E_{bd}(k,t)$
were 
defined as 
\begin{align}
E_{ud}(k,t) &= \frac{1}{2} \int_{|\bm{k}|=k} d \bm{k} 
|\bm{\hat{u}}_1 ( \bm{k},t) -  \bm{\hat{u}}_2 ( \bm{k},t)|^2 \ , \\
E_{bd}(k,t) &= \frac{1}{2} \int_{|\bm{k}|=k} d \bm{k} 
|\bm{\hat{b}}_1 ( \bm{k},t) -  \bm{\hat{b}}_2 ( \bm{k},t)|^2 \ ,
\label{eq:one}
\end{align}
with the difference energies being
$E_{ud}(t) = \int_0^\infty dk E_{ud}(k,t)$, and
$E_{bd}(t) = \int_0^\infty dk E_{bd}(k,t)$.
The wave number dependence of the spectra, $E_{ud}(k)$ and $E_{bd}(k)$, were 
similar to those found in pure hydrodynamics and similar to each other \cite{Berera2018}.

The maximal Lyapunov exponent of the fields could be obtained by looking
at the difference between the two fields.
We model the growth of $E_{ud}(t)$ and $E_{bd}(t)$ by
an exponential with $E_{ud}(t) \sim \exp(2\lambda_u t)$ and
$E_{bd}(t) \sim \exp(2\lambda_b t)$. In our simulations we found
that the exponential growth of both difference energies 
was the same and
occurred with approximately the same rate, and so
$\lambda_u \approx \lambda_b$ for all runs.
For reference this is termed the direct method
and all $\lambda$ quoted in this paper are found using this method,
except when explicitly stated otherwise.

As a way to cross-check our main method, we also use another common approach,
finite time Lyapunov exponents (FTLEs) \cite{OttBook}. This method does the following:
At a time $\delta t$ after the initial perturbation, and at every subsequent time interval of 
$\delta t$ the field is rescaled according to the rule
\begin{align}
\vu_2 \rightarrow \vu_1 + &\bigg( \frac{\delta_0}{\delta_u} \bigg) (\vu_2 - \vu_1) \ , \\
\vb_2 \rightarrow \vb_1 + &\bigg( \frac{\delta_0}{\delta_b} \bigg) (\vb_2 - \vb_1) \ .
\end{align}
Here $\delta_u = \delta_b = \sqrt{2 E_{ud} + 2 E_{bd}} = \delta$.
For each time interval an FTLE, $\gamma$, was defined
\begin{align}
\gamma = \frac{1}{\delta t} \ln \bigg(\frac{\delta}{\delta_0} \bigg) \ .
\end{align}
The FTLE method Lyapunov exponent, $\lambda_{F}$, 
is the average of these $\gamma$.


The FTLE method was used as
a check on the direct method. The time over which
the 
$\lambda_F$
were averaged was not very long due to simulation restrictions, however,
they did agree broadly with the other method and allowed us to be more sure of the
validity of the results. If we kept $\delta_u = \delta_b$ as coming from the union of the two fields,
none of the results were qualitatively changed if the FTLE
method was used as opposed to the direct method, and the quantitative changes were small.
The averaging procedure of the FTLE method 
tends to produce more stable results, but at the same
time this averaging may miss features such as the long term linear behavior.
However, it is the most effective tool
to use to test against the direct method we are using.
 
\section{Results}
\label{section5}

\subsection{Re and Pr Dependence}

We computed the dependence of the Lyapunov exponent $\lambda$ 
on Re and Pr by measuring $\lambda$ for a 
systematic grid of forced simulations.
We use the adjustable helical forcing 
to force the velocity field only with zero kinetic helicity injected for
a set of parameters of $\nu$ and $\eta$, which vary independently from 0.01
to 0.0003125.
We have computed a very large dataset in which 
Re varies from 50 to 2000, and Pr varies from 1/32 to 32.
The values of forcing resulted in a total dissipation $\varepsilon_t = \varepsilon_k + \varepsilon_b$ of 
$\simeq 0.1$.
A full table of simulation parameters is shown in Table \ref{tab:allsims}.
The set of simulations analysed in this section are the same as those
used for the main results of \cite{McKay2018}. 
These results may or may not be affected by the onset of dynamo action,
but systems showing dynamo action have been shown to be chaotic in the past \cite{Kurths1991}.
Specifically, if dynamo action is present or not, the degree of chaos may be changed.
However, the presence or absence of dynamo action is a further complication to the system
and this extra effect would overcomplicate the results.
As such, the effect of dynamo action on the simulations will not be addressed in any further
context in the present paper.
The effect of dynamo action on the degree of chaos should be a further area of study.

\begin{table*}
\caption{\label{tab:allsims} Simulation parameters, N is grid size, $\nu$ viscosity, Pr magnetic
Prandtl number, $\varepsilon_k$ kinetic dissipation, Re Reynolds number, $\lambda$
maximal Lyapunov exponent, $\sigma_\lambda$ standard deviation on $\lambda$, $T_0$
large eddy turnover time, $\tau$ kinetic Kolmogorov microscale time, $k_{max}\eta_k$ the resolution.
For $1024^3$ simulations, $k_{max} = 340$, for $512^3$ simulations, $k_{max} = 169$, 
$\eta_k = (\nu^3/\epsilon_k)^{1/4}$. A further discussion on the resolution 
of these simulations can be found in \cite{McKay2018}.
The data is publically available online \cite{datashare}.}
\begin{ruledtabular}
\begin{tabular}{cccccccccc}
N$^3$ & $\nu$ & Pr & $\varepsilon_k$ & Re & $\lambda$ & $\sigma_\lambda$ & $T_0$ & $\tau$ & $k_{max}\eta_{k} $ \\ 
 \hline
 \hline
$1024^3$ & 0.0003125 & 1/32 & 0.0990 & 1990 & 2.590 & 0.403 & 1.746 & 0.056 & 1.42 \\
$1024^3$ & 0.0003125 & 1/16 & 0.0423 & 2093 & 2.403 & 0.360 & 1.866 & 0.086 & 1.72 \\
$1024^3$ & 0.0003125 & 1/8 & 0.0240 & 2167 & 0.920 & 0.141 & 2.072 & 0.114 & 2.03 \\
$1024^3$ & 0.0003125 & 1/4 & 0.0217 & 2052 & 0.593 & 0.091 & 2.257 & 0.120 & 2.08 \\
$1024^3$ & 0.0003125 & 1/2 & 0.0267 & 2073 & 0.651 & 0.099 & 2.744 & 0.108 & 1.98 \\
$1024^3$ & 0.0003125 & 1 & 0.0352 & 2473 & 1.169 & 0.319 & 2.667 & 0.094 & 1.84 \\
\hline
$1024^3$ & 0.000625 & 1/16 & 0.1114 & 1104 & 1.912 & 0.296 & 1.653 & 0.075 & 2.33 \\
$1024^3$ & 0.000625 & 1/8 & 0.0588 & 1078 & 1.392 & 0.205 & 1.808 & 0.103 & 2.73 \\
$1024^3$ & 0.000625 & 1/4 & 0.0303 & 1112 & 0.560 & 0.058 & 2.149 & 0.144 & 3.22 \\
$1024^3$ & 0.000625 & 1/2 & 0.0305 & 1137 & 0.696 & 0.100 & 2.314 & 0.143 & 3.22 \\
$1024^3$ & 0.000625 & 1 & 0.0332 & 1119 & 0.535 & 0.082 & 2.535 & 0.137 & 3.15 \\
$1024^3$ & 0.000625 & 2 & 0.0393 & 1010 & 0.708 & 0.100 & 2.685 & 0.126 & 3.02 \\
\hline
$512^3$ & 0.00125 & 1/8 & 0.0616 & 552 & 0.839 & 0.119 & 2.007 & 0.142 & 2.26 \\
$512^3$ & 0.00125 & 1/4 & 0.0530 & 607 & 0.659 & 0.100 & 1.982 & 0.154 & 2.34 \\ 
$512^3$ & 0.00125 & 1/2 & 0.0367 & 604 & 0.420 & 0.056 & 2.410 & 0.185 & 2.57 \\
$512^3$ & 0.00125 & 1 & 0.0324 & 586 & 0.342 & 0.049 & 2.749 & 0.196 & 2.65 \\
$1024^3$ & 0.00125 & 2 & 0.0359 & 518 & 0.380 & 0.067 & 2.792 & 0.187 & 5.19 \\
$1024^3$ & 0.00125 & 4 & 0.0425 & 470 & 0.592 & 0.094 & 3.104 & 0.172 & 4.98 \\
\hline
$512^3$ & 0.0025 & 1/4 & 0.0755 & 267 & 0.655 & 0.093 & 1.975 & 0.182 & 3.60 \\
$512^3$ & 0.0025 & 1/2 & 0.0502 & 313 & 0.406 & 0.060 & 2.353 & 0.223 & 3.99 \\
$512^3$ & 0.0025 & 1 & 0.0433 & 296 & 0.352 & 0.052 & 2.523 & 0.240 & 4.14 \\
$512^3$ & 0.0025 & 2 & 0.0371 & 295 & 0.294 & 0.041 & 2.942 & 0.260 & 4.31 \\
$1024^3$ & 0.0025 & 4 & 0.0432 & 253 & 0.234 & 0.050 & 3.037 & 0.241 & 8.34 \\
$1024^3$ & 0.0025 & 8 & 0.0436 & 274 & 0.465 & 0.068 & 3.215 & 0.239 & 8.32 \\
\hline
$512^3$ & 0.005 & 1/2 & 0.0815 & 154 & 0.359 & 0.054 & 2.155 & 0.248 & 5.95 \\
$512^3$ & 0.005 & 1 & 0.0539 & 155 & 0.321 & 0.042 & 2.564 & 0.305 & 6.60 \\
$512^3$ & 0.005 & 2 & 0.0429 & 144 & 0.240 & 0.031 & 2.828 & 0.341 & 6.98 \\
$512^3$ & 0.005 & 4 & 0.0414 & 142 & 0.203 & 0.028 & 3.095 & 0.347 & 7.04 \\
$1024^3$ & 0.005 & 8 & 0.0467 & 139 & 0.320 & 0.046 & 3.289 & 0.327 & 13.75 \\
$1024^3$ & 0.005 & 16 & 0.0499 & 110 & 0.346 & 0.062 & 3.298 & 0.317 & 13.53 \\
\hline
$512^3$ & 0.01 & 1 & 0.0820 & 80.7 & 0.224 & 0.030 & 2.490 & 0.349 & 9.99 \\
$512^3$ & 0.01 & 2 & 0.0666 & 73.0 & 0.195 & 0.025 & 2.637 & 0.387 & 10.52 \\
$512^3$ & 0.01 & 4 & 0.0525 & 75.3 & 0.117 & 0.018 & 3.098 & 0.437 & 11.16 \\
$512^3$ & 0.01 & 8 & 0.0480 & 67.5 & 0.140 & 0.022 & 3.381 & 0.457 & 11.42 \\
$1024^3$ & 0.01 & 16 & 0.0540 & 60.7 & 0.133 & 0.029 & 3.414 & 0.431 & 22.30 \\
$1024^3$ & 0.01 & 32 & 0.0498 & 54.6 & 0.154 & 0.033 & 4.061 & 0.438 & 22.76 \\
\end{tabular}
\end{ruledtabular}
\end{table*}

\begin{table}[h!]
\begin{center}
\begin{tabular}{ c | c c c c c c}
 $\downarrow \nu$ $|$ $\eta \rightarrow$ & 0.01 & 0.005 & 0.0025 & 0.00125 & 0.000625 & 0.0003125 \\ 
 \hline
0.01	&	0.224	&	0.195	&	0.117	&	0.140	&	0.133	&	0.154	\\
0.005	&	0.359	&	0.321	&	0.240	&	0.203	&	0.320	&	0.346	\\
0.0025	&	0.655	&	0.406	&	0.352	&	0.294	&	0.234	&	0.465	\\
0.00125	&	0.839	&	0.659	&	0.420	&	0.342	&	0.380	&	0.592	\\
0.000625	&	1.912	&	1.392	&	0.560	&	0.696	&	0.535	&	0.708	\\
0.0003125	&	2.590	&	2.403	&	0.920	&	0.593	&	0.651	&	1.169	\\
\end{tabular}
\caption{Maximal Lyapunov exponent $\lambda$ for forced simulations with 
corresponding $\nu$ and $\eta$.}
\label{tab:nueta}
\end{center}
\end{table}

After a statistically steady state was reached, $\lambda$ was measured
using the direct method.
A grid of viscosities and diffusivities is shown in Table \ref{tab:nueta}.
The value quoted at each grid point is the corresponding $\lambda$.
What is seen very easily in the table is that those simulations with very low Pr 
(in the bottom left corner of the table) have much higher $\lambda$ 
than the high Pr simulations, which is also seen using the FTLE method.
However, for a fixed Re the Lyapunov exponent does not always decrease with increasing
Pr. Given the present database, a clear trend associated with Prandtl number is not
evident and more data is needed to confirm or not whether kinetic dominance
results in more chaotic dynamics.

\begin{figure}[!ht]
  \includegraphics[width=0.5\textwidth]{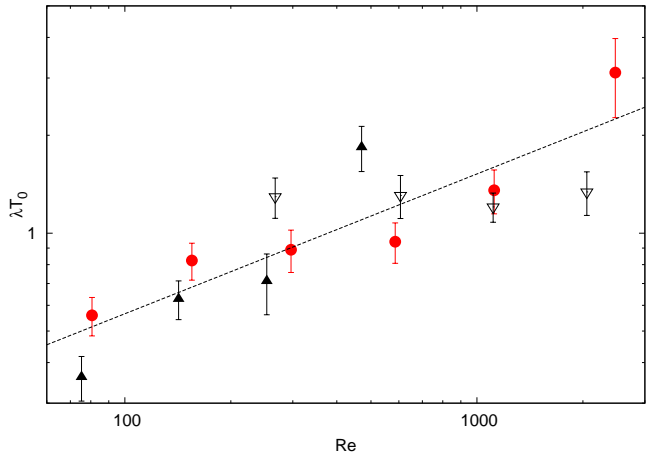}
  \caption{Lyapunov exponent $\lambda$ as a function of Re at different fixed values of Pr.
Pr $=$ 1 data is solid gray (red) circles, Pr $=$ 1/4 data is empty black downward
 triangles, Pr $=$ 4 data is solid black upward triangles. The dashed black line is
the best fit for the Pr $=$ 1 data.}
  \label{fig:pr14025}
\end{figure}

We now compare the dependence of $\lambda$ on Re for MHD
to that from hydrodynamics. Recall that the relation in hydrodynamics is
predicted to be as given Eq. (\ref{eq:alpha}).
The data in Table \ref{tab:nueta} can be split by Prandtl number and
then the dependence $\lambda T_0 \sim \ $Re$^\alpha$ is calculated separately.
This is shown for Pr $= 1/4,1,4$ in
Fig. \ref{fig:pr14025}. The dashed line is the best fit for Pr = 1, where
$\alpha = 0.43 \pm 0.09$
(the FTLE method produced an equivalent result within one standard deviation).

\begin{figure}[!ht]
\includegraphics[width=0.5\textwidth]{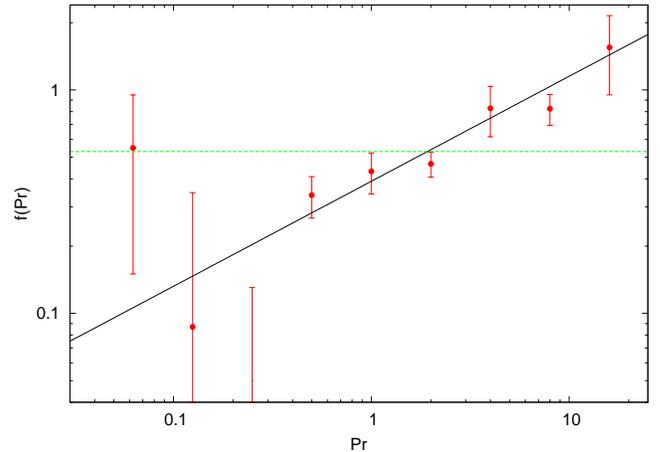}
\caption{We approximate $\lambda \sim Re^{f(Pr)}/T_0$, for individual Pr and then plot $f($Pr$)$.
Solid line is 0.39 Pr$^{0.47}$, dashed line is 0.53. The fit is log-log, and so the errors, though
they look large, especially at lower Pr, are not great in absolute value.}
\label{fig:fpr}
\end{figure}

One might wonder whether there is any dependence of $\lambda$ on Pr.
This question can be addressed by performing a fit of 
$\lambda T_0 \sim \ $Re$^\alpha$
for each Pr in turn.
To do this, the data for each Pr
was approximated as following Eq. (\ref{eq:alpha}), giving a
different exponent as a function of Pr,
$\alpha = f($Pr$)$.
This is equivalent to saying that $\alpha$ in Eq. (\ref{eq:alpha}), 
which is 0.53 in Navier-Stokes turbulence \cite{Berera2018} 
(or $\sim$0.64 in other simulations \cite{Boffetta2017,Mohan2017}), depends on the Prandtl number as
\begin{align}
\lambda \sim \frac{1}{T_0} Re^{f(\text{Pr})} \ ,
\end{align}
where $f(Pr)$ is some function of the Prandtl number, attaining roughly 1/2 at Pr = 1.
A plot of $f($Pr$)$ is shown in Fig. \ref{fig:fpr}.
The solid black line shows a power law behaviour $(0.39 \pm 0.09)$Pr$^{0.47 \pm 0.10}$,
whilst the dashed (green) line shows a constant $\alpha = 0.53$.
This will be discussed further below but these results show our data is consistent 
with there being no functional dependence of $\alpha$ on Pr.

\begin{figure}[!ht]
  \includegraphics[width=0.5\textwidth]{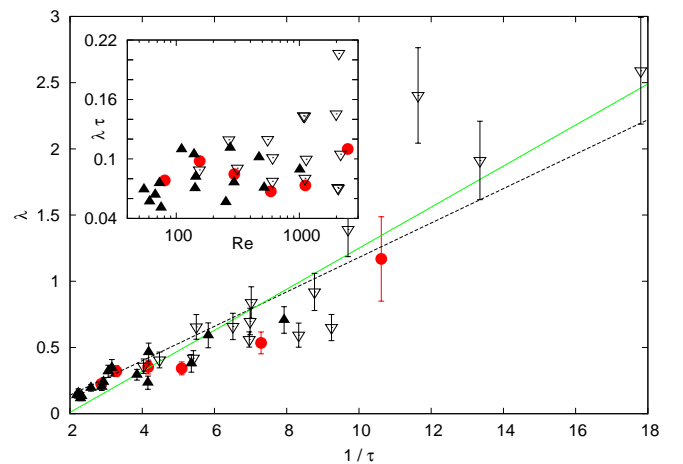}
  \caption{The inverse Kolmogorov time $1/\tau$ and $\lambda$, solid (green) line is fit for MHD data,
  dashed black line is fit for hydrodynamics \cite{Berera2018}.
  Pr = 1 data is solid gray (red) circles, Pr $<$ 1 data is empty black downward
 triangles, Pr $>$ 1 data is solid black upward triangles. 
	The inset shows no clear Re dependence.}
  \label{fig:kolmt}  
\end{figure}

Alternative to the above comparison between hydrodynamics and MHD,
we can directly test the relation $\lambda \sim 1/\tau$, which is our extension
of the Ruelle prediction to MHD.
This comparison is shown in Fig. \ref{fig:kolmt} and is
split according to Pr. Pr $=$ 1 data is solid gray (red) circles, Pr $<$ 1 data is empty black downward
triangles, and Pr $>$ 1 data is solid black upward triangles.
The inset shows $\lambda \tau$ against Re, and has no clear dependent trend.
Meanwhile, the main data shows broad agreement that $\lambda \sim 1/\tau$, although
there is an offset which may be related to the possibility that the onset of 
chaos in the system only occurs after
a certain level of turbulence is reached.
The Pr = 1 data is not noticeably different in trend compared to the Pr $\neq$ 1 data.
As such, the relation $\lambda \sim 1/\tau$ should not be dependent on just being at Pr = 1
but holds away from that point also.
In the figure, the solid (green) line shows the best fit to the data, whilst the
dashed black line shows the relationship between $\lambda$ and $1/\tau$ 
for hydrodynamics found in \cite{Berera2018}.
By eye, the relationship from pure hydrodynamic turbulence is not noticeably worse
than that of the fit to actual MHD simulations.

Although the data is noisy, this noise appeared in both the direct
and FTLE methods of measuring 
the Lyapunov exponent. 
As such, we expect that this noise is an inherent issue with MHD turbulence
as opposed to a measurement issue.
The noise may arise from the more complicated dynamics of MHD.
Specifically, although total magnetic helicity sums to zero, there are likely regions of 
opposite magnetic helicity which can reduce the chaos and 
such regions do not totally go away.
The data for the higher $\lambda$ in MHD is noisier (in general at higher Re)
than for hydrodynamics, whilst for low Re the data is comparatively clear.
However, it seems that the chaos is entirely dependent on the chaos due to the velocity 
field as also shown by the results from removing the Lorentz force later in the paper.
The results for higher $\lambda$ have greater variance than for lower ones.
This is probably due to the fact that these higher $\lambda$ simulations
are on the edge of our resolution range.
The fit does not weigh these heavily and so they should not overly affect the result.
Because of the noise in the data it was beneficial having two separate methods to 
analyse the data, which gave confidence in our interpretation.

We now make a comment about the suitability of the prediction
$\lambda T_0 \sim \ $Re$^\alpha$ or $\lambda \sim 1/\tau$,
both of which were assertions for which we offered some arguments but
nothing more solid.
From our data, we find that $\lambda \sim 1/\tau$ 
has a greater predictive quality and also
has no explicit Pr dependence.
The data is consistent with the idea that any dependence
of $\lambda$ on Pr comes indirectly through its dependence on $\tau$.

Regarding the prediction $\lambda T_0 \sim \ $Re$^\alpha$,
the data for $\alpha = f($Pr$)$ is statistically consistent with there being no 
dependence on Pr at all. 
The scaling $f($Pr$) \sim \ $Pr$^b$ has been used in Fig. \ref{fig:fpr}
simply for illustrative purposes. 
Fig. \ref{fig:fpr} shows that any Pr dependence of $\lambda$ is very unclear.
Indeed, on physical grounds, we would expect that we should reach finite values 
as Pr $\rightarrow 0$ and Pr $\rightarrow \infty$, which this fit does not.

In hydrodynamics, the prediction is that $\alpha = 1/2$, but is affected by intermittency
corrections to raise it above 1/2 \cite{Boffetta2017}.
A perfect scaling in MHD of $1/\tau \propto \ $Re$^{1/2}$ would result in $f($Pr$) = 1/2$.
From Fig. \ref{fig:fpr} we see that this is not a good fit to the data.
Although both Re and $\tau$ are based solely on kinetic quantities, Re is most affected by the
large scale quantities $u$ and $L$, whilst $\tau$ is most affected by the small scale dissipation.
The relative dominance of either will be affected by the Prandtl number, and so we should not
expect a perfect scaling $1/\tau \propto \ $Re$^{1/2}$, except maybe at some specific Prandtl number.
Any theoretical dependence between $\tau$ and Re in MHD may also need to 
take account of intermittency corrections, even at Pr = 1.
Thus, although our data is most consistent with $\lambda \sim 1/\tau$, this does not
imply a simple relationship $\lambda \propto \ $Re$^{1/2}$.
This is because the relationship between Re and $\tau$ in MHD is not the same as in hydrodynamic turbulence.

If, instead, $\lambda T_0$ were a function of Rm, we note that Rm$^f$ = Pr$^f$Re$^f$, and so the 
dependence of $\alpha$ on Pr
would be the same. 
That the results become independent of Pr is similar to results found in \cite{McKay2018}
for dissipation rates.
The dependence $\lambda \propto \ $Re$^a$ Pr$^b$ was also tested, but the dependence between
the values was less correlated than the simpler relationship $\lambda \sim 1/\tau$.

Indeed, if we want to be guided by the principle of universality, that at sufficiently
high Re or Pr we should have only a function of Pr or Re respectively, then such
a power law dependence for $\alpha$ on Pr would be precluded.
For example, this universality for dissipation rate was observed in \cite{Brandenburg2014}
and put on a firm footing in fully resolved simulations of MHD in \cite{McKay2018}.
These simulations show that there is complete dominance
of the dissipation by the kinetic dissipation, ie $\varepsilon_k + \varepsilon_b 
\simeq \varepsilon_k$ at sufficiently high Prandtl number and Rm.
At increasing Rm, the ratio of $\varepsilon_k/\varepsilon \rightarrow 1$,
and depending on the degree of accuracy required, the Lyapunov exponent can be approximated as
solely dependent on $\nu$ and 
$\varepsilon$, which would be consistent with an extension to MHD of the first Kolmogorov
hypothesis of similarity \cite{Kolmogorov1941a}.

Given that there are also many other time scales that could be used, such as
the Kolmogorov microscale time for the magnetic field or the large eddy turnover
time for the magnetic field, we did check the
relationship between these and $\lambda$ and found none that were better related
than $1/\tau$.
In any case, none of the others had theoretical justification.
As such, we suggest that our extension of the Ruelle theory to MHD
has the greatest predictive power consistent with our data.
This may be due to the nature of the MHD equations themselves.
As noted earlier, there is a direct non-linearity in the evolution of $\vec{u}$, whilst
$\vec{b}$ is only non-linear indirectly through the evolution of $\vec{u}$.
There is also an indirect non-linearity of $\vec{u}$ through the evolution of $\vec{b}$.
The direct non-linearity seems to be the most important for the chaos.

Since the exponential growth of $E_{ud}$ and $E_{bd}$ had the same exponent,
we surmise that if one field becomes sufficiently different, it drags the other one along
with it.
The data suggests that 
the difference in $\vec{u}$ drives the difference in $\vec{b}$.
This is backed by the fact that the relationship between $\lambda$ and $1/\tau$ for
hydrodynamics fits the data. This would be the case if the difference
in the magnetic field merely reacts to that in the velocity field, which is the true
driver of chaos in the system.

Because $\lambda$ is dependent on $\tau$ and not any MHD quantities,
we suggest that the direct non-linearity in the velocity field is
the most important feature and driver of the MHD chaos.
However, in some MHD systems dominated by magnetic reconnection it cannot be precluded that
the most important time scale would be the reconnection rate, which becomes
independent of $S$, the Lundquist number, for sufficiently high values of 
$S$ \cite{Loureiro2012}. We will not pursue these details in this paper.

\subsection{Magnetic helicity}
\label{sec:hel}


We now test the role that magnetic helicity plays in the chaotic
dynamics by adjusting the magnetic helicity in the system.
In contrast to the other simulations presented in this paper, in this subsection
the simulations had their magnetic field forced to control the magnetic helicity directly.
In these simulations, the velocity field was unforced.
The introduction of magnetic helicity will affect the mirror symmetry of the simulation
and may limit the applicability of HIT in them.

\begin{figure}[!ht]
  	\includegraphics[width=0.5\textwidth]{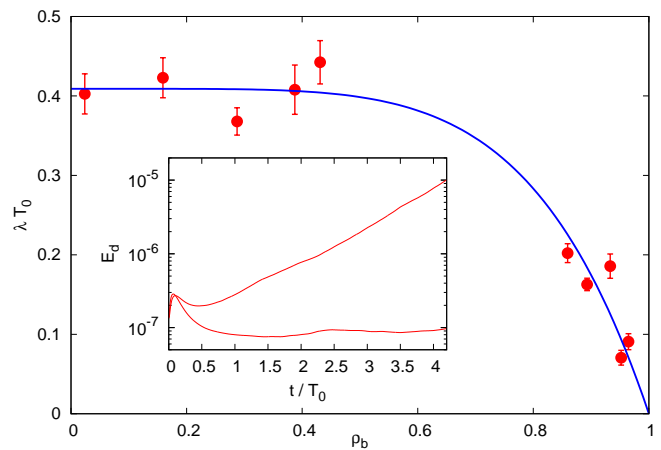}
	\caption{Influence of magnetic helicity on $\lambda$. Solid blue 
        line shows $\lambda = \lambda_0 (1 - \rho_b^{5.3})$. The inset shows 
        an example of a perturbation made on a helical and non-helical
field for Re $\sim$ 320. The helical field shows a diminution of chaos. }
	\label{fig:maghelandlambda}  
\end{figure}

To demonstrate the difference between simulations with and without
magnetic helicity, we present the results found in the inset of Fig. \ref{fig:maghelandlambda}.
The inset shows the response of two perturbations, one made in a simulation with maximal magnetic
helicity, and one made in a simulation with zero magnetic helicity. Both simulations
are otherwise equivalent with Re $\sim$ 320.
In the run without magnetic helicity, there is a 
regular exponential increase in $E_d = E_{ud} + E_{bd}$ whilst for the 
run with maximal magnetic helicity, after the initial perturbation, the two realisations remain close
for many large eddy turnovers, $T_0$. This effect is rather dramatic and indicates that
fully magnetically helical MHD may not be chaotic in the Eulerian sense.
Helicity has some organising effect on the fluid \cite{Brandenburg2005}.
As magnetic helicity increases, the organising effect becomes greater and the chaos diminishes.
We believe that the diminution of chaos cannot only be a result of the inverse transfer present in helical MHD.
For instance, in two-dimensional hydrodynamics there is an inverse transfer 
of energy, but still
a robust amount of chaos is seen \cite{Boffetta2001}.

To test the effect of magnetic helicity on chaos more systematically, 
we run a set of simulations where we vary the relative magnetic helicity, $\rho_b$,
defined as
\begin{align}
\rho_b^2 = \frac{\langle \vec{b} \cdot \vec{a} \rangle \langle \vec{b} \cdot \vec{a} \rangle}{\langle \vec{b} \cdot \vec{b} \rangle \langle \vec{a} \cdot \vec{a} \rangle} \ ,
\end{align}
with vector potential $\vec{b} = \nabla \times \vec{a}$.
We do this by injecting magnetic
helicity into the system with the adjustible helicity forcing described in \cite{McKay2017}.
All simulations were done with hypoviscosity, which took energy out at the large scale (low wave number) 
to ensure that the simulations did not blow up due to the inverse transfer present with magnetic helicity. 
After a statistically steady state was reached, including having roughly constant
$\rho_b$, the FTLE method was used to measure
the Lyapunov exponent $\lambda_F$
The simulations all have Re $\sim$ 120 and Pr = 1.

The results of this set of simulations are shown in Fig. \ref{fig:maghelandlambda}.
What is seen is that increasing $\rho_b$ does indeed create a diminution of chaos.
The data suggest that as $\rho_b \rightarrow 1$, $\lambda \rightarrow 0$.
Thus, we have assumed a parameter dependence of $\lambda = \lambda_0 (1 - \rho_b^n)$,
where $\lambda_0$ is the Lyapunov exponent at $\rho_b = 0$ and $n$ is some fit parameter,
here it is equal to $5.3 \pm 0.8$. This is shown in gray (blue) in the plot.
However, the uncertainty on $\lambda$ means that many other functional dependencies for
$\lambda$ on
$\rho_b$ are consistent, but these would require 
$\lambda \rightarrow 0$ at maximal magnetic helicity.
The systematic analysis suggests that as magnetic helicity
increases, the Eulerian chaos of the system decreases until it reaches zero
at maximal magnetic helicity. This would have important implications for the
Eulerian predictability of magnetically helical MHD systems. 
One might expect that astrophysical systems with large Reynolds numbers should have
nearly zero predictability.
However, if there is magnetic helicity present, the predictability could be dramatically increased.

It has been seen before that magnetic helicity plays an important role in MHD turbulent dynamics.
Specifically, even without initial magnetic helicity, the system is attracted to states that are
helical and laminar even with a large Re \cite{Dallas2015}.
These Beltrami states have greatly reduced non-linearity \cite{Servidio2008}.
This attractor behaviour is also seen in pure hydrodynamics \cite{Linkmann2015}, 
where these laminar states also exist.
The laminar states in hydrodynamics are purely attractive, such that once the
system becomes laminar, it does not delaminarize.
However, in MHD, the helical states can be exited spontaneously, but with a
tendency to stay near these states longer as magnetic helicity is increased.
This may explain the difference in behaviour of perturbations made in the laminar
states in hydrodynamics, where $E_d$ decays exponentially, and those made here,
where $E_d$ remains roughly constant.
In this sense, the MHD simulations act like the hydrodynamic simulations
in \cite{Berera2018} which have very low Re, but which have still
not relaminarized.

\subsection{Linear Growth}

In hydrodynamic turbulence, there is a limit on the growth rate of $E_{ud}$, which 
eventually seems to
grow linearly in time with rate equal to the dissipation rate. 
This result was found recently in \cite{Berera2018}
and subsequently confirmed in \cite{Boffetta2017}.
We report here for the first time 
an analogous effect in MHD. This behavior
can have important consequences 
for the predictability of
MHD systems with high Pr, such as galactic scale magnetic fields.

To test the dependence of the linear growth rate of $E_{ud}$ and $E_{bd}$
on the dissipation rates of each individual field, we ran a series of longer
simulations where we varied these dissipation rates indirectly.
This was done using a negative damping forcing, which maintained a constant total dissipation
rate $\varepsilon_t = \varepsilon_k + \varepsilon_b$, where $\varepsilon_k$ is the 
dissipation due to the velocity field and $\varepsilon_b$ is the dissipation due to the magnetic
field. This forcing maintained a rougly zero magnetic helicity and is fully described in \cite{McKay2017}.

\begin{figure}[!ht]
  \includegraphics[width=0.5\textwidth]{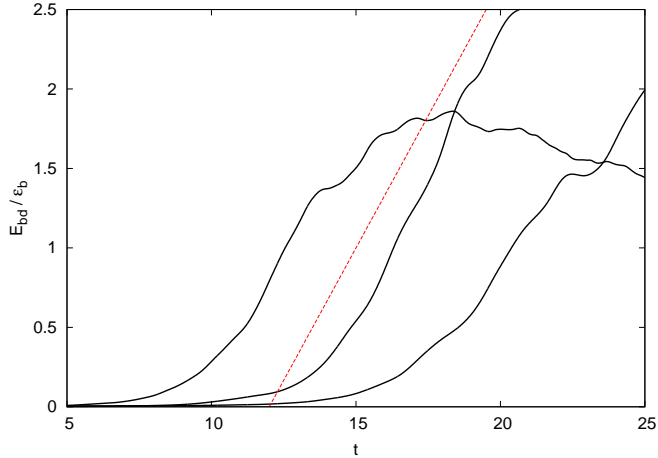}
	\caption{Linear growth of $E_{bd}/ \varepsilon_b$. Leftmost line has Re = 120 $\varepsilon_b$ = 0.11, middle has Re = 80 $\varepsilon_b$ = 0.036, rightmost has Re = 45 $\varepsilon_b$ = 0.0012. Dotted line shows $E_{bd}/ \varepsilon_b \sim t/3$.}
	\label{fig:lineareps}
\end{figure}

A series of three simulations was run with varying magnetic dissipation at fixed Pr = 1.
The simulations had $\varepsilon_b = 0.0012, 0.036,$ and 0.11, whilst Re = 45, 80, and 120 respectively.
The evolution of $E_{bd}/\varepsilon_b$ for these simulations is shown in Fig. \ref{fig:lineareps}.
The normalised linear growth rate for all simulations is roughly equal and implies that
$E_{bd} \sim \varepsilon_b t /3$.
The values for $\varepsilon_b$ used here differ by a factor of 100 so we feel relatively confident
that this linear relation is based upon the magnetic dissipation.
This resolves the question raised in Section \ref{section3},
namely on which dissipation rate did the linear growth rate rely.

These simulations also show eventual linear growth of $E_{ud}$.
This has the same form as in the hydrodynamic simulations of \cite{Berera2018,Boffetta2017}
and is not shown here.
For the growth of $E_{ud}$ it is difficult to distinguish whether the behaviour is based 
on $\varepsilon_k$ or $\varepsilon_t$, because the two values tended to be close. 
We suspect it is the former, and
this would be more appealing symmetrically, but the difference is less pronounced in
the values of the data. This is because it is difficult in practice to get $\varepsilon_b /
\varepsilon_k \gg 1$.

The fact that the exponential growth of both fields is the same, whilst the linear growth
at late times is different suggest that they are caused by fundamentally different processes.
Or at least they come from different aspects of dynamical systems theory, and that these
processes are controlled by different variables.

That eventually $E_{ud}$ and $E_{bd}$ have
differing linear growth rates is a new and very interesting result in our opinion, with important 
implications for predictability. Specifically, if $\varepsilon_b \rightarrow 0$, as happens
for high Pr \cite{Brandenburg2014,McKay2018}, whilst the magnetic field still has a significant amount
of energy in it, the predictability time would diverge.
Also, since there are differing growth rates, the difference of one field may become
saturated whilst the difference in the other field remains small. 
For example, even if velocity fields $\vu_1$ and $\vu_2$ are completely different,
the corresponding magnetic fields $\vb_1$ and $\vb_2$ can still be very similar.
Physically, this means that similar magnetic fields can have vastly different
velocity fields associated with them in MHD.

\subsection{FSLEs}

As an alternative to the direct method, the level of chaos can be quantified 
using finite size Lyapunov exponents (FSLEs) \cite{Aurell1997}.
These should not be confused with FTLEs.
These FSLE are defined by the time $T(\delta)$ that it takes for an error of size $\delta$ to grow by a
factor $r$.
Using this, an FSLE Lyapunov exponent can be defined $\Lambda(\delta) = \ln (r)/T(\delta)$.
For small $\delta$,
$\Lambda = \lambda$.

\begin{figure}[!ht]
  \includegraphics[width=0.5\textwidth]{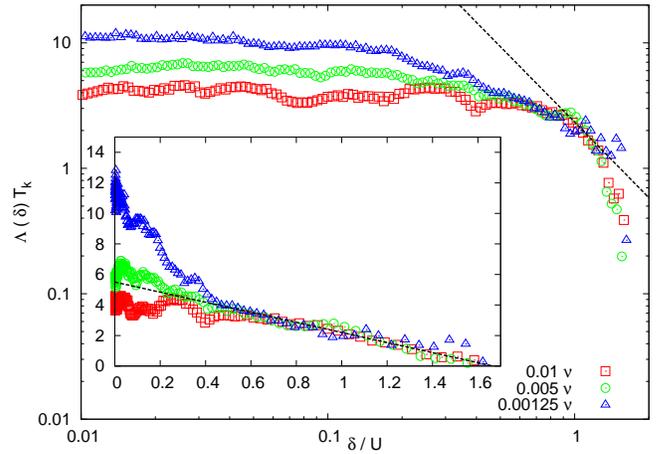}
        \caption{FSLE for the velocity field. Main plot dashed line shows $\sim \delta^{-2}$, inset dashed line is a straight line. 
		Main plot is log-log, inset is linear.
                 0.01 $\nu$ has Re = 80, 0.005 $\nu$ has Re = 155, 0.00125 $\nu$ has Re = 670.
		Time $T_k = E_u/\varepsilon_k$, and $U$ is the rms velocity.}
        \label{fig:ufsle}
\end{figure}

\begin{figure}[!ht]
  \includegraphics[width=0.5\textwidth]{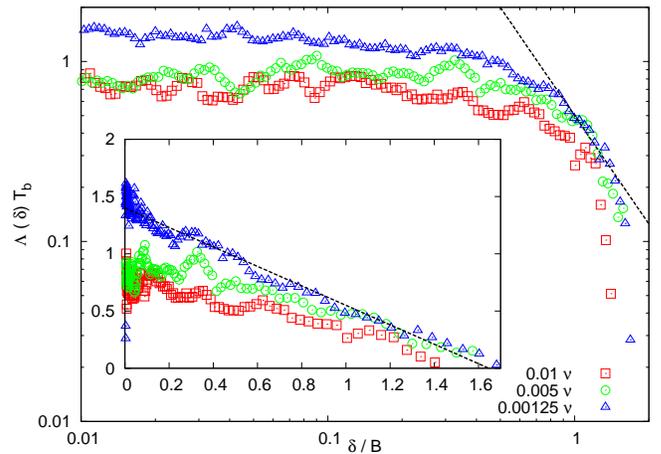}
        \caption{FSLE for the magnetic field, parameters same as in Fig. \ref{fig:ufsle}.
		Time $T_b = E_b/\varepsilon_b$, and $B$ is the rms magnetic field value.}
        \label{fig:bfsle}
\end{figure}

A set of five simulations was performed for three different Reynolds numbers
Re = 80, 155, and 670, each with Pr = 1.
After an initial small perturbation, the time taken for the perturbation to grow by a factor $r$
was measured and an average of these times was taken across the set of five simulations.
The velocity and magnetic fields were treated separately and 
$\Lambda(\delta)$
for each is shown
in Fig. \ref{fig:ufsle} and Fig. \ref{fig:bfsle} respectively.
For each field, $\delta$ is nondimensionalised
by the rms quantity of the relevant field.
Similarly, the FSLE $\Lambda(\delta)$ is non-dimensionalised by the
relevant timescale
$T_i = E_i/\varepsilon_i$, where $E_i$ is the energy of that field and $\varepsilon_i$ the dissipation
due to that field.

The main plot of Fig. \ref{fig:ufsle} and Fig. \ref{fig:bfsle}
presents $\Lambda(\delta)$
with logarithmic scales, whilst the inset is the same data presented with
linear scales. In the logarithmic plot, a scaling of $\Lambda(\delta) \sim \delta^{-2}$ 
is shown as a dashed line.
Lorenz has suggested that for fluid turbulence a disturbance in the inertial
range will grow with rate equal to the local eddy turnover time \cite{Lorenz1969}.
This eventually predicts that $\Lambda(\delta) \sim \delta^{-2}$ \cite{Boffetta2017}.
However, due to the interplay of the different fields and the potential presence of non-local
interactions between the $\vu$ and $\vb$ fields, there is no great justification for this
in MHD.
The inset of the figures shows a linear fit to data which is closer over a larger range
of values than the $\delta^{-2}$ fit.
The data is presented both linearly and logarithmically to
facilitate comparison with hydrodynamic results.
Although the fit $\Lambda(\delta) \sim \delta^{-2}$ is good for hydrodynamic turbulence \cite{Boffetta2017},
it is not especially good for MHD turbulence
as can be seen in Fig. \ref{fig:ufsle} and Fig. \ref{fig:bfsle}.
Thus, there should be a different mechanism for the growth of disturbances
in hydrodynamic turbulence as opposed to MHD.
Whilst the relation $\Lambda(\delta) \sim \delta^{-2}$ results in late linear growth of disturbances,
the linear relation $\Lambda(\delta) \sim \delta$ can also result in a restriction of the late
growth rate.

The shape of the FSLE plot is related to the later linear
growth rate of $E_{ud}$ and $E_{bd}$. This means that the slope onto which $\Lambda(\delta)$
collapses should be dependent on the corresponding dissipation rate. Whilst
the data is nosier than for hydrodynamics, this is again like all the previous data
we have found.
In both linear and logarithmic plots, the FSLE collapse onto a slope which is roughly independent
of Re. This collapse is more consistent for the velocity field.
The parameters used did not ensure a constant ratio $\varepsilon_k/\varepsilon_b$, which varies
from 2 for Re = 80 to 1/2 for Re = 670. As such, we are fairly confident that the relevant
dissipation for determining the dynamics is that of the respective field and not the total dissipation.
In this way, the fields, despite being dependent on each other non-linearly, have linear behaviour dictated by their
own evolution.

The linear dependence of $\Lambda$ on $\delta$ implies that
$E_d(t)$ has a maximum growth rate which is proportional 
to the dissipation rate.
This is shown by the following argument.
In the region where $\Lambda$ is a linear function of $\delta$, 
$\Lambda(\delta) = c - m\delta = \dot{\delta}/\delta$ where $m$ and $c$ are positive constants.
This last equality comes from a small expansion of $r$ about 1 in
the definition of $\Lambda(\delta) = \ln (r)/T(\delta)$.
This differential equation is solved by
\begin{align}
\delta = \frac{c}{m + \Big( \frac{c}{\delta_0} - m \Big) e^{-ct}} \ ,
\end{align}
where $\delta_0$ is the separation at $t = 0$.
This gives $E_d(t) = \delta^2/2$ a sigmoid shape, with maximum rate of growth when $E_d$ is at half
of its maximum. The maximum of $\delta$ is $c/m$. At $\delta_0 = c/2m$, $\partial_t E_d(0) = c^3 / 8m^2$.
Remembering the normalisation of $\Lambda$ and $\delta$ we redefine $c = a/T_i$ and
$m = b/UT_i$.
From the simulation data presented in Fig. \ref{fig:ufsle} and Fig. \ref{fig:bfsle},
we find that
$a = 6$ for the velocity field and $a = 1.5$ for the magnetic field.
In both cases, $a/b = 1.5$.  The rms values
are defined $U^2 = 2E_k/3$ and $B^2 = 2E_b/3$.
Thus, the maximum rate of growth for the relevant difference has
\begin{align}
\partial_t E_{id}(t) \leq \frac{9a}{48} \varepsilon_i \ .
\end{align}
For $a = 6$, this predicts a maximum of $54/48 \varepsilon_k$, where $54/48 = 1.125$
such that $\partial_t E_{ud} \leq 1.125 \varepsilon_k$. This value
of 1.125 is very
close to the 1.12 previously found in hydrodynamic turbulence \cite{Berera2018}. This also implies
that the magnetic field has $\partial_t E_{bd} \leq 27/96 \varepsilon_b$,
with 27/96 close to the value of 1/3 previously found.
This argument shows, as we said above, that $E_d(t)$ has a maximum growth rate which
is proportional to the dissipation rate.

A previous study of dynamics of systems with different timescales, of which MHD turbulence is definitely
an example, looked at the study of both scales through the use of FSLE \cite{Boffetta1998}.
It concludes with a note that parametrization of the fast scales is not crucial and that the
slow mode dynamics are dominant. In MHD turbulence, the dynamics of the velocity field are
equivalent to the fast modes and the magnetic field to the slow modes.
As such, we also predict that the velocity field can be successfully parametrized
whilst still capturing the most important magnetic field dynamics, as is done in
the static field approximation.

\subsection{Influence of the Lorentz force}

We further test our hypothesis that the chaos in the MHD system comes mainly
from the velocity field evolution using the diagnostic tool of removing the Lorentz force.
The MHD equations Eq. (1-2) can be changed to omit the action of the Lorentz force on the velocity field.
Although this means that the fields no longer conserve energy, we use this
as a diagnostic tool to disentangle the amount of chaos
that comes from perturbations to each field.
This has been performed in prior simulations of MHD turbulence investigating the inverse transfer
of energy, finding that this Lorentz force term 
is not necessary for inverse transfer to persist \cite{Berera2014}.
In our simulations, if the energy spectra when decreasing with wave number were approximated
as following $k^{-n}$, the $n$ for both fields was somewhat smaller than for
an equivalent simulation with a Lorentz force.

We perform a simulation using the adjustible helicity forcing as described previously
for Pr = 1 with $\nu = 0.01$ and Re = 74,
where we have removed the Lorentz force $(\nabla \times \vb) \times \vb$.
This makes the magnetic field equation linear. Even with the Lorentz force, 
the non-linearity for $\vb$ in the MHD equations comes entirely from $\vu$.
In this case, the magnetic field alone cannot produce chaos or turbulence, but this gets input
from $\vu$.
Either the magnetic or velocity field can be perturbed individually.
Both perturbations were made from identical instances of the fields.
Because of the decoupling of the velocity field from the magnetic field
a magnetic field perturbation would never induce a velocity perturbation but a velocity field perturbation
can induce a magnetic perturbation. 

When the magnetic field was perturbed with the Lorentz force removed, the
magnetic field perturbation can remain relatively stable for many large eddy turnover times.
The ad hoc removal of the Lorentz force is sometimes done in seed dynamo simulations, where it is argued
that the magnetic field is too small to affect the velocity field.
Because of this tendency, the Reynolds number used was a modest 74, otherwise the magnetic field
has a tendency to accumulate energy indefinitely.
However, in simulations which did have an unphysical exponential growth of magnetic energy \cite{Berera2014},
the perturbation in the magnetic field grew no faster than the magnetic field itself did.

When the velocity field was perturbed with the Lorentz force removed
both magnetic and velocity fields diverged at the same rate, which
agrees with our interpretation that it is the velocity field which drives the chaos.

Although the results in Fig. \ref{fig:kolmt} are noisier than for hydrodynamic turbulence,
this diagnostic tool of removing the Lorentz force is another result which supports our 
theoretical argument in Section \ref{section3}.
Namely, it also shows that the
chaotic properties are dominated by the velocity field properties, here being the dependence on
$1/\tau$ as opposed to any other timescales. 
This diagnostic tool also shows that the magnetic field itself is not
chaotic. This is important for simulations where the Lorentz force is removed by affecting the
statistics, such as in seed dynamo simulations.

In the absence of a Lorentz force, the velocity field
behaves the same as in hydrodynamic turbulence.
Since, by removing the Lorentz term, our results are not
significantly changed, we can state that the chaos in the system with a magnetic field is not
significantly different to that without it. As such, our prior assumptions that
the magnetic field should not affect the chaos greatly
are in agreement with our results, and are strengthened by them.

\section{Discussion and Conclusion}
\label{sectionConclusion}

The data presented here include Pr which cover three orders of magnitude.
Although many of the results here are found to be 
independent of magnetic Prandtl number, there
is direct comparison with many physical systems.
Here we look at some of them in turn and see how our measurements can be useful for their further study.

In toroidal plasmas, such as those found in tokamak reactors, Pr is expected to be on the order 
of 100 \cite{Itoh1993,Mendonca2018}.
However, the turbulence is strongly confined and should be greatly affected by the boundary conditions,
as such, an assumption of homogeneity and isotropy will not capture the full dynamics.
There may be other practical reasons not to apply MHD as a model for the plasmas in tokamak reactors.
Even so, the level of chaos in the system should be related to the generation of instabilities in the 
flow. These instabilities can affect the confinement of the plasma. As well, the level of instability
should be affected by the level of inverse cascade, which is greater at increased Rm.

Accretion disks around black holes and neutron stars are predicted to have regions where
Pr is close to unity \cite{Balbus2008}.
Though these should be affected by relativistic effects, if the results in this paper extend beyond the 
Prandtl number range of our simulations they may also apply to more typical accretion disks.
By understanding the ratio of the dissipations, we can understand the relative importance
of ion and electron heating in these systems which has an effect on the luminosity and thus
their observational characteristics.
In regions with greater chaos of the particles, those regions with lower Kolmogorov microscale
time and lower magnetic helicity, the process of accretion should be reduced.
If particles are, on average, moving together and not drifting apart exponentially, there
should be a longer time for them to accrete.
As such, in any accretion disk, we expect that the accretion rate should be increased
by magnetic helicity and $\tau$. Indeed, simulations have shown already that
accretion disks are associated with turbulent dynamo action, 
which is also associated with magnetic helicity \cite{Brandenburg2005}.
Thus, an increased accretion rate should be associated with diminution of chaos, here consistent with
increased magnetic helicity.

Understanding the chaos seen in MHD turbulence is important for understanding the scope
for prediction of turbulent conducting fluids. The findings here may also be applicable
to coupled dynamical system where there is a direct non-linearity for only one of the fields
and multiple relevant timescales.

Our extension of the Ruelle prediction to MHD that $\lambda \sim 1/\tau$ is 
most consistent with our own data.
We conclude that this is because the MHD equations
are directly non-linear only in $\vec{u}$ and indirectly through the coupling for $\vec{b}$.
We have also confirmed previous findings that show magnetic helicity results in a diminution
of chaos and further predict that a fully magnetically helical system should have zero Lyapunov exponent,
although these previous findings were not found in DNS \cite{Zienicke1998,Escande2000}.

One of the most interesting results is the new finding that the growth of both $E_{ud}$ and $E_{bd}$
becomes linear with rates that depend on the dissipation rate of the relevant field.
This may apply more generally to non-linearly coupled fields.
Specifically in the case of high Pr, where $\varepsilon_k$ becomes dominant and $\varepsilon_b$
very small, then this has important implications for the long term predictability of galactic
plasmas and magnetic fields.
For high Re, we should expect $\lambda$ to be very large and that any
small scale error should grow in size very quickly and so the the separation between the two fields
will enter the linear regime very quickly.
Thus, any predictability time will be
dominated by $E/\varepsilon$ for the relevant field. For magnetic fields, this can become
extremely large, and if there is a large amount of magnetic helicity in the system, then the
predictability time of the magnetic fields can become very long.

\begin{acknowledgments}
We thank Mairi E. McKay for helpful discussion.
This work has used resources from ARCHER \cite{archer} via the Director's Time budget.
This work used the Cirrus UK National Tier-2 HPC Service at EPCC \cite{cirrus} funded
by the University of Edinburgh and EPSRC (EP/P020267/1).
R.D.J.G.H is supported by the U.K. Engineering and Physical Sciences Research Council (EP/M506515/1),
D.C. is supported by the University of Edinburgh.
A.B. acknowledges funding from the U.K. Science and Technology Facilities Council.
\end{acknowledgments}

%
%

\end{document}